\documentstyle[12pt,twoside,fleqn,espcrc2]{article}

\input{boxedeps}
\HideDisplacementBoxes
\SetepsfEPSFSpecial

\newcommand{\G}{\gamma}
\newcommand{\GG}{{\gamma\gamma}}
\newcommand{\EPEM}{e^+e^-}

\newcommand{\BE}{\begin{equation}}
\newcommand{\EE}{\end{equation}}

\makeatletter
\def\lesssim{\mathrel{\mathpalette\vereq<}}
\def\vereq#1#2{\lower3pt\vbox{\baselineskip1.5pt \lineskip1.5pt
\ialign{$\m@th#1\hfill##\hfil$\crcr#2\crcr\sim\crcr}}}

\makeatother

\title{Photon-Photon and Photon-Hadron Physics at Relativistic Heavy Ion 
Colliders}

\author{K. Hencken$^{\rm a}$, P. Stagnoli$^{\rm a}$, D. Trautmann
\address{Universit\"at Basel, Klingelbergstr. 82, CH-4056 Basel}
and  
G. Baur \address{Forschungszentrum J\"ulich,D-52425 J\"ulich}
}

\begin{document}
\maketitle

\section{Introduction}
The parton model is very useful to study scattering processes at very
high energies. For example, nuclei consist of nucleons, which in turn 
consist of quarks and gluons, photons consist of lepton pairs, electrons 
of photons, etc.. Relativistic nuclei have photons as an important 
constituent at low enough virtuality $Q^2=-q^2$ of the photon, due to the 
coherent action of all charges in the nucleus. The coherence condition limits
the virtuality of the photons to
\BE
Q^2 \lesssim 1/R^2,
\EE
where the radius of a nucleus is given
approximately by $R=1.2$~fm~$A^{1/3}$, with $A$ the nucleon
number. From the kinematics of the process one has
$Q^2=\frac{\omega^2}{\G^2}+q_\perp^2$.
This limits the maximum energy of the quasireal photon and the perpendicular 
component of its momentum to
\BE
\omega<\omega_{max} \approx \G/R,
\qquad
q_\perp \lesssim 1/R,
\EE
where $\G$ is the Lorentz factor of the projectile. The ratio
$x=\omega/E$, where $E$ denotes the energy of the nucleus, is
smaller than $x< x_{max}=1/(R M_N A)$ (we always use
$\hbar=c=1$). We find $x\protect\lesssim 10^{-3}$ for Ca ions and
$\protect\lesssim 10^{-4}$ for Pb ions. 

The collisions of $e^+$ and $e^-$ has been the traditional way to
study $\GG$-collisions. Similarly photon-photon collisions can also be
observed in hadron-hadron collisions. Of course, the strong interaction 
of the two nuclei has to be taken into consideration. Up to now 
hadron-hadron 
collisions have not been used for two-photon physics. An exception can 
be found in \cite{Vannucci80}, where the production of $\mu^+\mu^-$
pairs at the ISR was studied.  The special class of events was
selected, where no hadrons are seen associated with the muon pair.
In this way one makes sure that the hadrons do not interact strongly
with each other, i.e., one is dealing with peripheral collisions 
(impact parameters $b>2R$).
\begin{figure}[tbh]
\begin{center}
\ForceWidth{0.5\hsize}
\BoxedEPSF{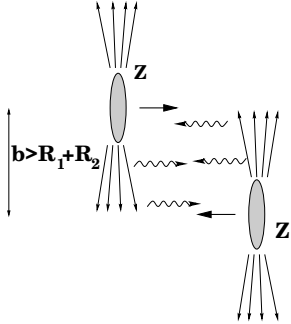}
\caption{
Two fast moving electrically charged nuclei are an abundant source of
(quasireal) photons. They can collide with each other and with the
other nucleus. For peripheral collisions with impact parameters $b>2R$,
this is useful for photon-photon and photon-nucleus
collisions.}
\label{fig_collision}
\end{center}
\end{figure}

Due to the coherent action of all protons the photon flux in heavy ion 
collisions is large ($\sim Z^2$) and therefore of interest for 
photon-photon and 
photon-hadron physics. This subject has been studied for several years. 
Recent reviews of this topic are \cite{BaurHT98,KraussGS97}.
The ``Relativistic Heavy Ion Collider'' (RHIC)
will have a program to investigate such collisions
experimentally \cite{Nystrand98} and similar programs are discussed at 
LHC \cite{Sadovsky93,HenckenKKS96,Felix97,BaurHTS98}. 
At RHIC ($\G\approx 100$) the equivalent photon spectrum extends up to
several GeV. Therefore the available invariant mass range is up to
about the mass of the $\eta_c$.  When the ``Large Hadron Collider''
will be scheduled in 2004/2008, the study of these reactions can be 
extended to both higher luminosities but also to much higher invariant
masses, hithero unexplored. We quote J. D. Bjorken \cite{Bjorken99}:
{\it It is an important portion (of the FELIX program at LHC
\cite{Felix97}) to tag on
Weizsaecker Williams photons (via the nonobservation of completely
undissociated forward ions) in ion-ion running, creating a high
luminosity $\GG$ collider.}
\begin{figure}[bth]
\begin{center}
\ForceWidth{0.8\hsize}
\BoxedEPSF{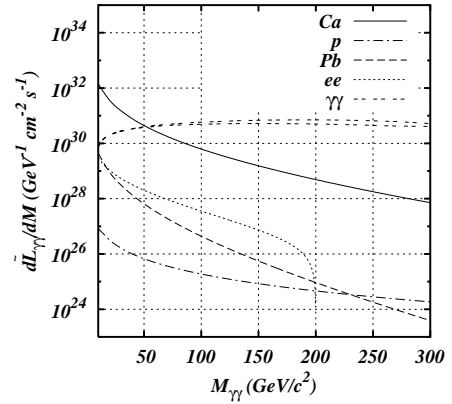}
\caption{
Comparison of the effective $\GG$-Luminosities 
($d\tilde L_{\GG}/dM = L_{AA} \times dL_{\GG}/dM$) for different ion 
species. For comparison the same
quantity is shown for LEP2 (``$ee$'') and a future NLC/PLC (next linear
collider/photon linear collider, ``$\gamma\gamma$''), where photons are 
obtained by laser backscattering; the results for two different 
polarizations are shown.}
\label{fig_lum}
\end{center}
\end{figure}

In Fig.~2 we compare the effective $\GG$ luminosities, that is, the product
of the beam luminosity with the two-photon luminosity 
($L_{AA} \times dL_{\GG}/dM$) for various collider scenarios.
We use the following collider parameters: LEP2: $E_{el}=100$GeV,
$L=10^{32} cm^{-2} s^{-1}$, NLC/PLC: $E_{el}=500$GeV, 
$L=2 \times 10^{33} cm^{-2} s^{-1}$, Pb-Pb heavy-ion mode at LHC: 
$\gamma=2950$,
$L=10^{26} cm^{-2} s^{-1}$, Ca-Ca: $\gamma=3750$,
$L=4 \times 10^{30} cm^{-2} s^{-1}$,p-p: $\gamma=7450$,
$L=10^{34} cm^{-2} s^{-1}$.
In the Ca-Ca heavy ion mode, higher 
effective luminosities can be achieved as, e.g., in the Pb-Pb mode, 
since higher AA luminosities can be reached there.
Since the event rates are proportional to the luminosities, and 
interesting events are rare (see also below), we think that it is important
to aim at rather high luminosities in the ion-ion runs.
This should be possible, especially for the medium heavy ions like Ca.
For further details see \cite{BrandtEM94,HenckenTB95}.

\section{$\gamma A$ collisions}

The interaction of quasireal photons with protons has been studied
extensively at the electron-proton collider HERA,
where the quasi-real photons from the electron (positron) beam are
used. Similar and more detailed studies will be possible at the
relativistic heavy ion colliders RHIC and LHC using the larger flux
of quasireal photons from one of the colliding nuclei. In the
photon-nucleon subsystem, one can reach invariant masses $W_{\G N}$ up
to $W_{\G N,max}=\sqrt{4 W_{max} E_N} \approx 0.8 \G A^{-1/6}$~GeV. For
Pb at LHC ($\G=2950$) one obtains 950~GeV and even higher
values for Ca. Thus one can study physics quite similar to the one at
HERA, with nuclei instead of protons. 

Photon-nucleon physics includes many aspects, like the energy
dependence of total cross-sections, diffractive and non-diffractive
processes.
An important subject is the elastic vector meson production $\G p
\rightarrow V p$ (with $V=\rho,\omega,\phi,J/\Psi,\dots$).
Here one gets insight into the interface 
between perturbative QCD and hadronic physics. Elastic processes, that is, 
the proton remaining in its ground state, as well as inelastic ones,
are of interest. For the hard exclusive photoproduction 
of $J/\Psi$ mesons, recent data from HERA have shown a rapid increase of 
the total cross section with $W_{\G N}$, in agreement with perturbative QCD.

Such studies could be extended to photon-nucleus interactions at
RHIC, thus complementing the HERA studies. Of special interest 
is the coupling of the photon of one nucleus to the Pomeron-field of
the other nucleus. Such studies are envisaged for RHIC, see
\cite{KleinS97b}.
Estimates of vector meson production in photon-nucleon processes at 
RHIC and LHC are given in \cite{BaurHT98}. In $AA$ collisions one can 
have incoherent photoproduction on the individual $A$ nucleons as well 
as coherent one. Shadowing effects will occur in the nuclear environment
and it will be interesting to study these. For the coherent production
one expects an $A^{4/3}$ dependence at large momentum transfer, in
contrast to the $A^2$ dependence, one sees, e.g., in low energy $\nu$A
scattering. Calculations \cite{KleinN99} for RHIC show an
$A$-dependence between these two extremes. In this context,  RHIC and
LHC can also be thought of as vector meson factories \cite{KleinN99}.

\section{Photon-Photon Physics}
\label{ssec_ggphysics}

Up to now photon-photon scattering has been mainly studied at $\EPEM$
colliders.
The traditional range of invariant masses has been the region of mesons,
ranging from $\pi^0$ ($m_{\pi^0}=135$~MeV) up to about $\eta_c$
($m_{\eta_c}=2980$~MeV). Also the total $\GG\rightarrow$~hadron 
cross-section was studied, e.g., at LEP2 \cite{L3:97,ph99gg}. 
In contrast to $\EPEM$ colliders, the photons from heavy ions have 
always $Q^2\lesssim 1/R^2$ due to the coherence condition and are 
therefore in all these cases quasireal (``untagged'').

Two-photon collisions give access to most of the $C=+1$ mesons. In
principle $C=-1$ vector mesons can be produced by the fusion of three
or more equivalent photons. But this effect is smaller than the
contribution coming from $\G$-A collisions (see above),
even for nuclei with large $Z$ (see \cite{BaurHT98}).

While the $\GG$ invariant masses, which will be reached at RHIC, will
mainly be useful to explore QCD at lower energies, the $\GG$ invariant
mass range at LHC --- up to about 100 GeV --- will open up new
possibilities. This includes the
discovery of the Higgs-boson in the $\GG$-production channel or new
physics beyond the standard model, like supersymmetry or
compositeness.

A number of calculations have been made for a medium heavy standard
model Higgs \cite{DreesEZ89,MuellerS90,Papageorgiu95,Norbury90}.
Chances of finding the standard model Higgs in this case are 
only marginal \cite{BaurHTS98}.
An alternative scenario with a light Higgs boson was, e.g., given in
\cite{ChoudhuryK97} in the framework of the ``general two Higgs
doublet model'', which allows for a very light particle in the
few GeV region. The authors of \cite{ChoudhuryK97} proposed to look
for such a  light neutral Higgs boson at the proposed low energy
$\GG$-collider.  We want to point out that the LHC Ca-Ca heavy ion
mode would also be  very suitable for such a search.

One can also speculate about new particles with strong coupling to the
$\GG$-channel. Large $\Gamma_{\GG}$-widths will directly lead to large
$\GG$ production cross-sections, see \cite{Renard83,BaurFF84}. Since
the $\GG$-width of a resonance is mainly proportional to the wave
function at the origin, huge values can be obtained for very tightly
bound systems. Composite scalar bosons at $W_{\GG}\approx 50$~GeV are
expected to have $\GG$-widths of several MeV
\cite{Renard83,BaurFF84}. The search for such kind of resonances in
the $\GG$-production channel will be possible at LHC.
%
%
\begin{figure*}[tbh]
\begin{center}
\ForceWidth{0.405\hsize}
\BoxedEPSF{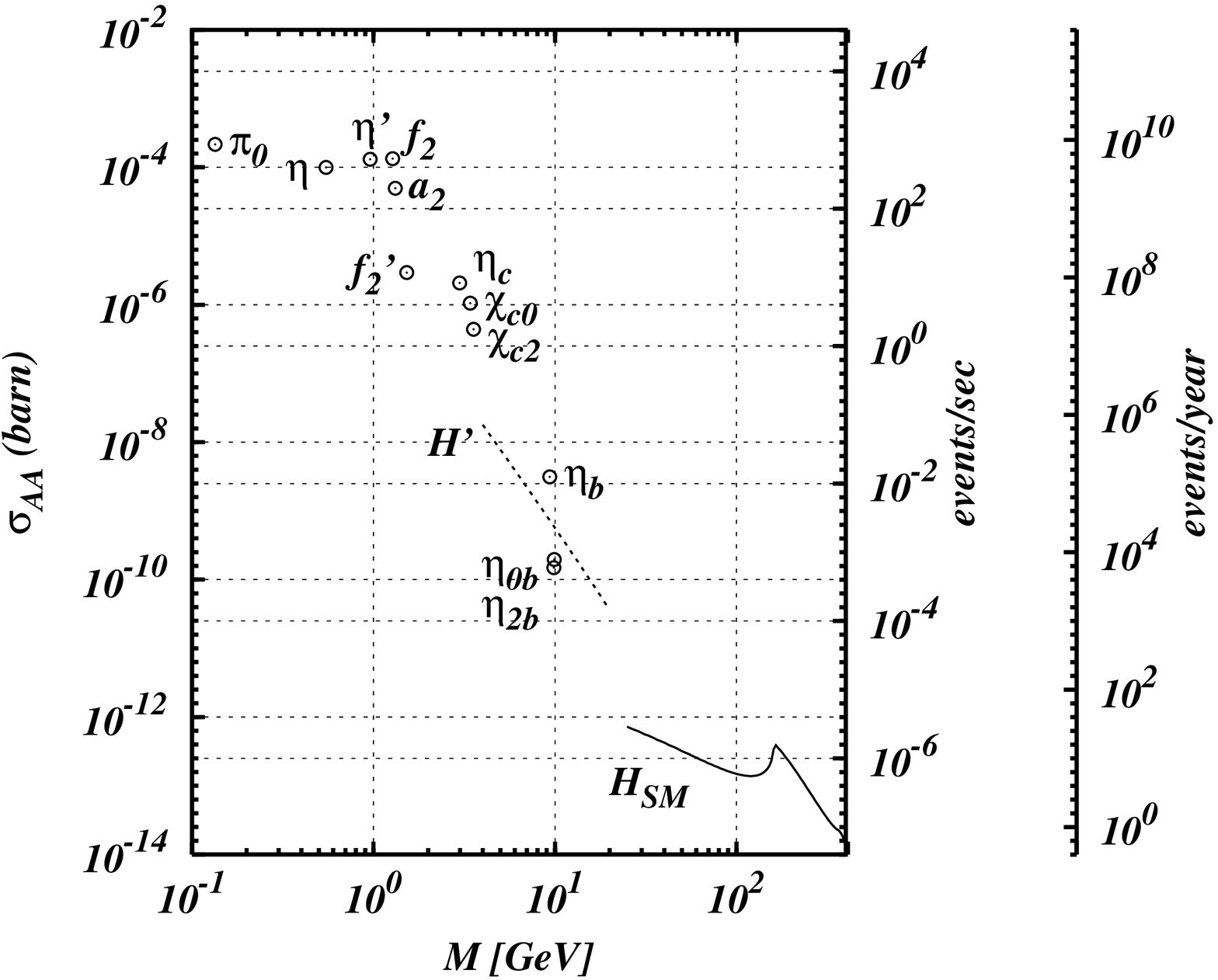}
(A)~~
\ForceWidth{0.405\hsize}
\BoxedEPSF{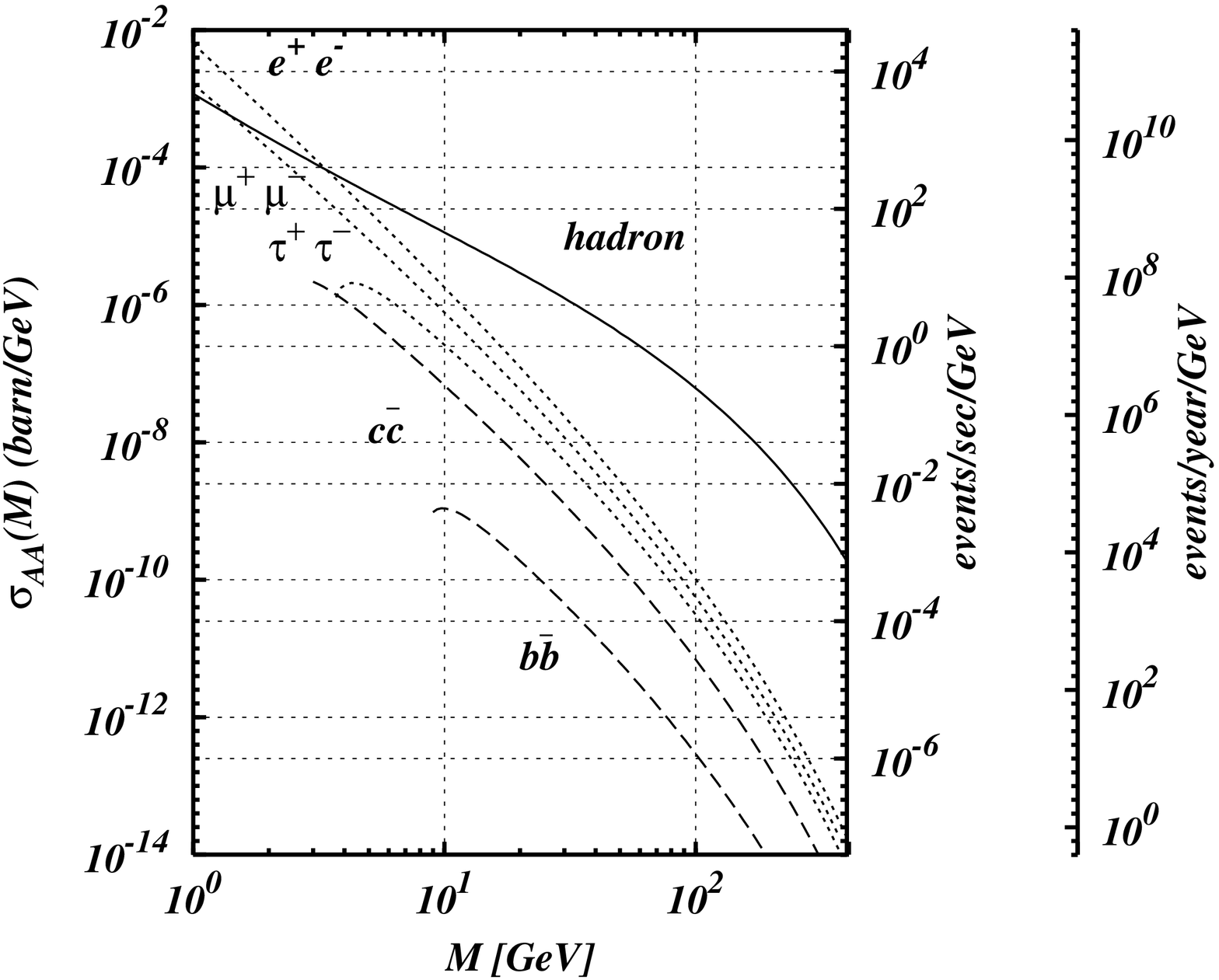}
(B)
\end{center}
\caption{
Overview of the total cross section and production rates of 
resonances (A) and continuum states (B) in Ca-Ca collisions at the LHC
using a beam luminosity of $L_{AA}=4 \times 10^{30} cm^{-2} s^{-1}$ and
$10^7$s per year.
$H_{SM}$ denotes the standard model Higgs, $H'$ the one in the ``general
two Higgs doublet model''. For further details see \cite{BaurHT98}.}
\end{figure*}

In Refs. \cite{DreesGN94,OhnemusWZ94} $\GG$-processes at $pp$
colliders were studied. It was observed that non-strongly interacting
particles (sleptons, charginos, neutralinos, and
charged Higgs bosons) are difficult to detect in hadronic collisions
as Drell-Yan and $gg$-fusion yield only low production rates for such
particles. Therefore producing such particles in $\GG$ interactions
was examined. Clean events can be expected, if the protons do not break
up in the photon emission process. In \cite{DreesGN94} it was also
pointed out, that at the high luminosity for $pp$ collisions
at LHC about 16 minimum bias events per bunch crossing are
expected. These hadronic background events are not a big concern in 
heavy ion collisions, due to their much smaller luminosities.

Similar considerations for new physics are also made in connection
with the planned $eA$ collider at DESY (Hamburg). Again, the coherent
field of a nucleus gives rise to a $Z^2$ factor in the cross-section
for photon-photon processes in $eA$ collisions \cite{KrawczykL95}.

An overview of the expected events rates for Ca-Ca collisions at LHC is 
given in Figure~3. Both production of resonances and continuum states are 
shown. 

An interesting topic in itself is the $\EPEM$ pair 
production. The fields are strong enough to produce multiple pairs in a
single collisions. A discussion of this subject together with calculations 
within the semiclassical approximation can be found in 
\cite{HenckenTB95b,AlscherHT97}, see also \cite{ph99serbo}.

\section{Conclusion}

Electromagnetic processes, that is, photon-photon and photon-hadron
collisions, are an interesting option for heavy ion colliders,
complementing the program for central collisions. It is the study of
``silent events'', with  relatively small multiplicities and a 
small background. The method of equivalent photons is a well
established tool to describe these kinds of reactions; effects arising
from the more complex structure of the ions are well under
control. Remaining uncertainties coming , e.g., from triggering can be 
eliminated by using a luminosity monitor from muon-- or
electron--pairs \cite{Sadovsky93}. A trigger for peripheral
collisions is essential in order to select photon-photon events. Such
a trigger seems to be possible based on the survival of the nuclei
after the collision and the use of the small transverse momenta of the
produced system \cite{KleinS97b}. Difficult to judge quantitatively
at the moment is the influence of strong interactions in grazing
collisions, i.e., effects arising from the nuclear stratosphere and
Pomeron interactions. 

The high photon fluxes open up possibilities for studies up to
energies hitherto unexplored. At RHIC the invariant mass range extends
up to several GeV. At the LHC one also has the possibility to study
new physics in the 100 GeV range.

Peripheral collisions using Photon-Pomeron and Pomeron-Pomeron
collisions, that is, diffractive processes are an additional
application. They use essentially the same triggering conditions and
therefore one should be able to record them at the same time as
photon-photon events.


\end{document}